# Gravity emerging from direct-action EM in a toy universe of electrons and positrons

**M Ibison**
Institute for Advanced Studies at Austin,
11855 Research Boulevard, Austin TX 78759-2443, USA

E-mail: ibison@earthtech.org

**Abstract.** We sketch the derivation of a Newtonian gravity-like force emerging from a direct-action variant of classical electromagnetism. The binding energy is a consequence of maximal phase correlation of the sources mediated by approximately monochromatic direct-action fields. The resulting force then has the character of a strong version of the van der Waals force, whose superior strength can be attributed to relatively coherent primary fields (compared with the totally incoherent effects of the ZPF). The model also predicts the existence of a background having some of the character of dark energy.

## 1. Introduction

This document presents some results that provide support for the existence of time-symmetric electromagnetic interactions involving equal positive combinations of advanced and retarded fields. According to common experience however, electromagnetic interactions are exclusively retarded. Retardation establishes an electromagnetic arrow of time, which, coincidentally, agrees with the thermodynamic arrow of time – the direction of increasing entropy. Since these two could be interrelated, a conservative application of a time-symmetric theory with no danger of conflict with observation should be confined to systems lacking *any* arrow of time – thermodynamic or electromagnetic. This is the state of affairs at zero Kelvin, which, therefore, is the exclusive domain of enquiry of this document.

Though the possibility of time-symmetric EM interactions at 0K may be novel, it seems quite likely that accommodation of this alternative in QED would amount to no more than a re-interpretation of the existing electromagnetic zero-point field (EM ZPF), without affecting the predictions of the theory. But the more radical proposal under investigation here is that time-symmetric EM fields exist in the *classical* theory. This effort may be regarded as an investigation of the consistency of this classical picture with the zero Kelvin state of affairs in QED in which the electromagnetic field is just the ZPF, but where the matter is given (i.e. the QFT Fermion fields are not in their ground state). To be consistent with the possible correspondence with the ZPF, such classical fields should presumably be ubiquitous, homogeneous, and isotropic, and have an omega-cubed spectrum [1]. This is exactly the situation that forms the basis of 'Stochastic Electrodynamics' (SED), (also called Random Electrodynamics), which has been championed by Boyer [2-4], Marshall [5,6], Puthoff [7], and others. However, here it will be necessary to make some crucial departures from the SED paradigm in order to derive the subsequent results[1].

---

[1] Though in the SED literature there is no mention of time-symmetry, it follows that because the background random radiation of SED is a homogeneous isotropic vacuum field that is not sourced, one cannot say if it is retarded,

## 2. Discussion of the model

*2.1 Direct Action Version of Electromagnetism*
We will assume that electromagnetism is most accurately described by the direct-action version of the theory [8-12] as opposed to the Maxwell version. For a review of the issues see the introduction in [13], the article by Pegg [14], and the detailed discussion in Davies' book [15]. In the direct-action theory electromagnetic radiation is *necessarily* time-symmetric. The theory then has the problem of explaining the apparent absence of advanced radiation. It is important to observe, in accord with the discussion above, that this is an issue only at elevated temperatures, i.e. above absolute zero. The contribution of Wheeler and Feynman [11,12] was to show that the advanced radiation would effectively be cancelled locally by secondary emission from distant future absorbers, provided such absorbers are present with density sufficient to ensure complete absorption of all retarded radiation on the future light cone. The viability of the Wheeler-Feynman implementation of direct-action EM therefore rests on the cosmological boundary conditions, which, as it turns out, are not met in the commonly acceptable cosmologies [15,16]. An alternative explanation for the emergence of time-asymmetry at elevated temperatures tied more directly to cosmological expansion has been sketched by the author [17]. A more comprehensive argument is in preparation and will be presented shortly.

*2.2 Classical Zitterbewegung*
One of several important consequences of adopting the direct-action approach is that the classical background field (CBF) standing in for the ZPF must be *sourced*. It follows that these sources must be in non-uniform motion at zero Kelvin. Predominantly the sources responsible for maintaining a ubiquitous background field will be electrons, since they have the largest scattering cross-section. In keeping with the attempted correspondence with QED at 0K it seems natural, therefore, to associate this classical motion with the zitterbewegung of the Dirac electron. In accord with work of Barut and Zanghi [18], Hestenes [19-21], Rodrigues, Vaz, Recami, and Salesi [22,23], the default assumption will be that the point electron executes more or less circular motion at twice the Compton frequency. In the direct action theory the motion of all such particles, and the (time-symmetric) radiation fields that result, must be mutually self-consistent[2]. Self-consistency at the level of time-averaged energies can be established using the virial theorem (see below). More detailed analysis will be required to demonstrate that the approximately circular motions assumed here are mutually self-consistent (i.e. as maintained by the direct-action fields).

*2.3 Lorentz Invariance of the Classical Background Field*
There may appear to be a problem with the above proposal: if the zitterbewegung motion is taken literally, i.e. as describing the motion of classical point particle, then the resulting CBF will be monochromatic at twice the Compton frequency - in contrast with the $\omega^3$ spectral energy density of the ZPF and the classicized version employed by SED. Recall it is important for the Lorentz-Invariance of QED that the $\omega^3$ spectrum and the statistics of the ZPF are Lorentz Invariant. Since a monochromatic field is not Lorentz Invariant, it might therefore appear that interactions with a monochromatic CBF such as proposed here will reveal motion with respect to an absolute frame, predicting a frame-dependent quality for the electron. But there is a crucial difference between the two theories. QED has the means of detecting Doppler shifts of an EM field using a standard of length and time that does not depend on the state of motion, namely the Compton wavelength associated with the rest mass. That is, QED asserts that the rest mass is a *Lorentz scalar*. By contrast, in the theory under investigation here, the electron has no

---

advanced, or some combination. Consequently such distinctions must be immaterial to any subsequent analysis. One is at liberty therefore, to regard the SED EM vacuum field as time-symmetric.

[2] This is distinct from the SED paradigm wherein the ZPF dictates the motion of the classical particles. In that case the particles cannot source the fields because their motion cannot supply the power necessary to generate the theoretically infinite energy density.

intrinsic rest mass and therefore can provide no intrinsic scale of length or time. Instead, these are acquired from interaction with the CBF[3]. The circumstances under which this difference would be sufficient to maintain Lorentz-Invariance are not addressed in this document.

*2.4 The massless charge*
In keeping with the attempted correspondence with QED at 0K, we are guided by the zitterbewegung of the Dirac electron in modeling the motion of charges. To be specific, it will be convenient to regard the observed mass of the electron as Lorentz-boosted from a very small 'bare' value, acquiring, as a result of EM interactions with other charges, a speed arbitrarily close to the speed of light. (Variants of SED have been published, wherein a *structure* which can store internal energy is energized by the ZPF [24], are quite different from the structure-less interaction considered here. See [25] for a discussion of this point.) If $m_e$ is the observed mass of the electron appearing in the Dirac equation, then

$$m_e = m_{\text{bare}} / \sqrt{1 - \mathbf{v}^2} \qquad (1)$$

where $m_{\text{bare}} \to 0$, $|\mathbf{v}| \to 1$ as the quantity on the right of (45) tends to the finite observed value[4]. Unless otherwise stated, we work in units where $c = 1$ and $\varepsilon_0 = \mu_0 = 1/4\pi$. At 0K, the EM interactions responsible for the boost are the time-symmetric fields of other charges executing similar approximately circular motion. Of course (1) implies that the *observed* 'rest' mass is not really a rest mass, nor is it necessarily fixed, though fixing the motion, as we are doing here, does fix the mass.

*2.5 Electromagnetic Action for the Theory*
Implementation of this scheme (of direct-action with massless charges moving at light speed) may be accomplished through minimization of the action

$$I = -\frac{1}{2} \sum_{\substack{j,k \\ j \neq k}} e_j e_k \int dt \int dt' \left(1 - \mathbf{v}_j(t) \cdot \mathbf{v}_k(t')\right) \delta\left((t-t')^2 - (\mathbf{x}_j(t) - \mathbf{x}_k(t'))^2\right) - m_{\text{bare}} \sum_j \int dt \sqrt{1 - \mathbf{v}_j^2(t)} \qquad (2)$$

and then letting $m_{\text{bare}} \to 0$ in the solutions of the Euler equations. As discussed in [13,14,26], the exclusion of self action ($j = k$) in the sum, whilst avoiding infinite electromagnetic mass, can be no more than a useful approximation. Other actions, as for example that suggested by Barut and Zanghi [18], may turn out to be superior.

*2.6 Cosmological Model*
Throughout this document, where it is necessary to be specific, we will assume the traditional steady-state cosmology of Hoyle et al [27] wherein the densities of matter and radiation are constant, the Hubble radius is constant ($H(t) = H_0$), and the Hubble radius is identical with the maximum radius of electromagnetic contact, here denoted by *R*: $R = cH_0$, [15,28]. Consequently, the number of fermions (primarily electrons) and baryons (primarily protons) are constant, and herein assumed to be equal and denoted by *N*, having value of order $10^{79}$, depending on various assumptions (see below). Consideration of more realistic cosmologies must be deferred to a later article.

---

[3] The theory must be scale free. Lengths associated with masses must, in the end, be knowable only as ratios. One expects then to be able to find some theoretical basis for the numerical values of such ratios at all scales, including for example Dirac's ratio between the Hubble radius and the classical electron radius.

[4] This approach is fundamentally different to that taken in [13], wherein electromagnetic self-action is retained in full and not offset by a mechanical mass. However, the essential feature of the approach taken in this paper - that the mass is determined dynamically – can probably also be derived from the un-renormalized model. In that case the self-action would be retained, with the expected outcome that the electromagnetic mass would be rendered finite when the charge is in non-uniform motion at light speed.

## 3. Interaction Energy

*3.1 Interaction of a Pair of Sources*
The retarded electric field of a single source at the origin in the far field at point $r$ is the real part of [29]

$$\mathbf{E}_{\text{ret}}(t,\mathbf{r}) = \hat{\mathbf{e}} \frac{ek}{r} e^{i(k(t-r)+\beta)} \tag{1}$$

where $\beta$ is an arbitrary phase. $\hat{\mathbf{e}}$ is a unit vector in the direction of the electric polarization. In accord with the position advocated in this document we assume that at zero Kelvin that there is an equal contribution from advanced and retarded fields, in which case the electric field is

$$\mathbf{E}(t,\mathbf{r}) = \frac{1}{2}\mathbf{E}_{\text{ret}}(t,\mathbf{r}) + \frac{1}{2}\mathbf{E}_{\text{adv}}(t,\mathbf{r}) = \frac{1}{2}\hat{\mathbf{e}}\frac{ek}{r}e^{i(k(t-r)+\beta)} + \frac{1}{2}\hat{\mathbf{e}}\frac{ek}{r}e^{i(k(t+r)+\beta)} = \hat{\mathbf{e}}\frac{ek}{r}e^{i(kt+\beta)}\cos(kr). \tag{2}$$

And so the far field from a zero-Kelvin source located at $\mathbf{r}_i$ is

$$\mathbf{E}_i(t,\mathbf{r}) = \hat{\mathbf{e}}_i \frac{ek}{|\mathbf{r}-\mathbf{r}_i|} e^{i(kt+\beta_i)} \cos(k|\mathbf{r}-\mathbf{r}_i|). \tag{3}$$

Consider now the *interaction* energy between a pair of sources at $\mathbf{r}_i$ and $\mathbf{r}_j$. We are not interested here in the *self*-energy, which is excluded in this model. Instead we are interested in the contribution to the total energy that can be attributed to the spatial placement, relative phase, and relative polarization of these two charges, given that they are immersed in the background radiation field dictated, predominantly, by the $N$ other charges. Recalling that radiation fields have equal energy in the magnetic and electric fields, the total energy is

$$\varepsilon_{\text{int}} = \frac{1}{2}\int d^3r \left(\mathbf{E}^2 + \mathbf{B}^2\right)^{\dagger} = \int d^3r \, \mathbf{E}^{2\dagger} = \int d^3r \left(\sum_{i=1}^{N} \text{Re}\{\mathbf{E}_i(t,\mathbf{r})\}\right)^{2\dagger}. \tag{4}$$

where the dagger is there to remind us that self energy terms are excluded by fiat. Expanding the quadratic therefore,

$$\varepsilon_{\text{int}} = \sum_{i=1}^{N}\sum_{j=i+1}^{N} \varepsilon_{ij}; \quad \varepsilon_{ij} = 2\int d^3r \, \text{Re}\{\mathbf{E}_i(t,\mathbf{r})\}\text{Re}\{\mathbf{E}_j(t,\mathbf{r})\}. \tag{5}$$

Clearly there will be rapidly oscillating components at frequency $2k$. But here we will restrict attention to the energy averaged over several periods, which, for a complex field is ([30])

$$\bar{\varepsilon}_{ij} = \text{Re}\left\{\int d^3r \, \mathbf{E}_i(t,\mathbf{r})\mathbf{E}_j^*(t,\mathbf{r})\right\}. \tag{6}$$

Putting (3) into (6) gives

$$\bar{\varepsilon}_{ij} = e^2 k^2 \, \text{Re}\left\{\hat{\mathbf{e}}_i \cdot \hat{\mathbf{e}}_j^* e^{i(\beta_i - \beta_j)}\right\} \int d^3r \, \frac{\cos(k|\mathbf{r}-\mathbf{r}_i|)\cos(k|\mathbf{r}-\mathbf{r}_j|)}{|\mathbf{r}-\mathbf{r}_i||\mathbf{r}-\mathbf{r}_j|}. \tag{7}$$

If the integration is taken to be over a closed hyper-surface then we are free to move the origin without affecting the result. Exercising that freedom by letting $\mathbf{r} \to \mathbf{r} + \mathbf{r}_j$, the above becomes

$$\bar{\varepsilon}_{ij} = e^2 \omega^2 \operatorname{Re}\left\{\hat{\mathbf{e}}_i \cdot \hat{\mathbf{e}}_j^* \, e^{i\beta_{ij}}\right\} \int d^3 r \, \frac{1}{|\mathbf{r} - \mathbf{r}_{ij}|} \cos(kr) \cos\left(k|\mathbf{r} - \mathbf{r}_{ij}|\right) \tag{8}$$

where $\beta_{ij} \equiv \beta_i - \beta_j$, are constant phases, and $\mathbf{r}_{ij} \equiv \mathbf{r}_i - \mathbf{r}_j$ is the vector displacement between the two sources. (This result is changed by a numerical factor of order unity for plausible alternative integration geometries.) In the above one has

$$|\mathbf{r} - \mathbf{r}_{ij}| = \sqrt{r^2 + r_{ij}^2 + 2 r r_{ij} \cos\theta}, \tag{9}$$

where $\theta$ is the angle between the field point $\mathbf{r}$, and the displacement vector $\mathbf{r}_{ij}$: $\cos\theta = \hat{\mathbf{r}} \cdot \hat{\mathbf{r}}_{ij}$. Since the integration is over all angles the result is insensitive to the choice of orientation of the axes. Using that freedom to align the $z$ axis with $\mathbf{r}_{ij}$, (8) can be written

$$\bar{\varepsilon}_{ij} = e^2 k^2 \operatorname{Re}\left\{\hat{\mathbf{e}}_i \cdot \hat{\mathbf{e}}_j^* \, e^{i\beta_{ij}}\right\} \int_0^R dr\, r \int_0^\pi d\theta \sin\theta \int_0^{2\pi} d\phi \, \frac{\cos(kr)\cos\left(k\sqrt{r^2 + r_{ij}^2 + 2 r r_{ij} \cos\theta}\right)}{\sqrt{r^2 + r_{ij}^2 + 2 r r_{ij} \cos\theta}}. \tag{10}$$

We have restricted the radial integration to radius $R$. In a steady-state universe this is a fixed quantity and is equal to the Hubble radius. In more conventional cosmologies the approach above is not accurate and is in need of refinement; $R$ should be the radius of communication, which differs, in general, for advanced and retarded influences. The integration over $\phi$ simply gives $2\pi$. To perform the integration over $\theta$ it is useful first to make a change of variable

$$\xi = \sqrt{r^2 + r_{ij}^2 + 2 r r_{ij} \cos\theta} \Rightarrow \frac{d\xi}{d\theta} = -\frac{r r_{ij} \sin\theta}{\sqrt{r^2 + r_{ij}^2 + 2 r r_{ij} \cos\theta}} \Rightarrow \theta \in [0, \pi] \mapsto \xi \in \left[r + r_{ij}, |r - r_{ij}|\right]. \tag{11}$$

With this, (10) becomes

$$\bar{\varepsilon}_{ij} = -2\pi e^2 k^2 \operatorname{Re}\left\{\hat{\mathbf{e}}_i \cdot \hat{\mathbf{e}}_j^* \, e^{i\beta_{ij}}\right\} \int_0^R dr\, r \int_{r + r_{ij}}^{|r - r_{ij}|} d\xi \, \frac{\cos(kr)\cos(k\xi)}{r r_{ij}}$$
$$= \frac{2\pi e^2 k}{r_{ij}} \operatorname{Re}\left\{\hat{\mathbf{e}}_i \cdot \hat{\mathbf{e}}_j^* \, e^{i\beta_{ij}}\right\} \int_0^R dr \cos(kr)\left(\sin\left(k(r + r_{ij})\right) - \sin\left(k|r - r_{ij}|\right)\right) \tag{12}$$

The modulus necessitates segmentation of the range of integration:

$$\bar{\varepsilon}_{ij} = \frac{4\pi e^2 k}{r_{ij}} \operatorname{Re}\left\{\hat{\mathbf{e}}_i \cdot \hat{\mathbf{e}}_j^* \, e^{i\beta_{ij}}\right\} \left[\int_0^{r_{ij}} dr \cos(kr)\sin(kr)\cos(kr_{ij}) + \int_{r_{ij}}^R dr \cos^2(kr)\sin(kr_{ij})\right]$$
$$= \frac{2\pi e^2 k}{r_{ij}} \operatorname{Re}\left\{\hat{\mathbf{e}}_i \cdot \hat{\mathbf{e}}_j^* \, e^{i\beta_{ij}}\right\} \left[\cos(kr_{ij}) \int_0^{r_{ij}} dr \sin(2kr) + \sin(kr_{ij}) \int_{r_{ij}}^R dr \left(1 + \cos(2kr)\right)\right] \tag{13}$$
$$= \frac{\pi e^2}{r_{ij}} \operatorname{Re}\left\{\hat{\mathbf{e}}_i \cdot \hat{\mathbf{e}}_j^* \, e^{i\beta_{ij}}\right\} \left[\cos(kr_{ij})\left(1 - \cos(2kr_{ij})\right) + \sin(kr_{ij})\left(2k(R - r_{ij}) + \sin(2kR) - \sin(2kr_{ij})\right)\right]$$

In the far field the term proportional to $2k(R - r_{ij})$ is overwhelmingly larger than the other terms. Therefore (13) can be simplified to

$$\bar{\varepsilon}_{ij} \approx 2\pi e^2 k \left( R/r_{ij} - 1 \right) \text{Re} \left\{ \hat{\mathbf{e}}_i \cdot \hat{\mathbf{e}}_j^* e^{i\beta_{ij}} \right\} \sin(kr_{ij}). \tag{14}$$

In accord with the earlier discussions and in particular [20], we assume that the oscillations are at twice the Compton frequency, $k = 2m_e/\hbar$, in which terms (14) is

$$\bar{\varepsilon}_{ij} \approx 4\pi \alpha m_e \left( R/r_{ij} - 1 \right) f_{ij}, \tag{15}$$

where $\alpha$ is the fine structure constant and where we have implicitly defined a phase-factor of order unity:

$$f_{ij} \equiv \text{Re} \left\{ \hat{\mathbf{e}}_i \cdot \hat{\mathbf{e}}_j^* e^{i\beta_{ij}} \right\} \sin(kr_{ij}). \tag{16}$$

*3.2 Total Interaction Energy*

Since we are ignoring self-action, the total energy in the EM field is the sum over all particles of the pair-wise interaction energies:

$$\bar{\varepsilon}_{\text{int}} = \sum_{i=1}^{N} \sum_{j=i+1}^{N} \bar{\varepsilon}_{ij} \approx 4\pi \alpha m_e \sum_{i=1}^{N} \sum_{j=i+1}^{N} f_{ij} \left( R/r_{ij} - 1 \right). \tag{17}$$

These $f_{ij}$ depend on the relative angle between the polarizations of the two electric fields, on the initial relative phase $\beta_{ij}$, and on the phase shift at frequency $k$ over the distance $r_{ij}$. An argument permitting a very approximate estimation of $f_{ij}$ now follows.

Consistent with the above declaration that this time-symmetric analysis is valid at zero Kelvin only, the total energy (17), which has $N(N-1)/2 \approx N^2/2$ terms, must be at a minimum.[5] Clearly there are $N$ degrees of freedom at our disposal in the 'initial' phases $\beta_i$. The polarization vectors can be varied also; if the sources are in circular motion the polarization vectors can be written

$$\hat{\mathbf{e}}_i = \frac{\hat{\mathbf{u}}_i + i\hat{\mathbf{v}}_i}{\sqrt{2}}, \quad \hat{\mathbf{u}}_i \cdot \hat{\mathbf{v}}_i = 0 \tag{18}$$

where $\hat{\mathbf{u}}$ and $\hat{\mathbf{v}}$ are real orthogonal unit vectors. There are 2 degrees of freedom in each of $\hat{\mathbf{u}}$ and $\hat{\mathbf{v}}$, subject to one constraint, (i.e.: they are orthogonal). Hence there are $3N$ degrees of freedom in the set of $\hat{\mathbf{e}}_i$.[6] There are an additional $3N$ degrees of freedom in the $\mathbf{r}_i$ that make up the 'position' phases $kr_{ij} \equiv k|\mathbf{r}_j - \mathbf{r}_j|$: the $\mathbf{r}_i$ need change by no more than a Compton wavelength in order to exercise full variation of these phases, and this change cannot be resolved at the level of the macroscopic separations $r_{ij}$. This argument holds even when the particles are in macroscopic motion (i.e., in addition to zitterbewegung) exercising, say, orbital motion about a nucleus. In that case, though those $\mathbf{r}_i$ are now changing macroscopically in time, roughly speaking, at each point on the macroscopic path, the particle can make small Compton-sized adjustments in order to exercise the $3N$ degrees of freedom in the $N^2$ contributions from the $kr_{ij}$.

---

[5] From the perspective of QFT, the matter is not in its ground state. The latter implies the absence of real matter and the presence of only the Fermionic ZPF – the electron-positron Dirac sea. The zero Kelvin condition stipulated here minimizes the free energy, assuming the matter is given.

[6] If elliptical motion is allowed, then the resulting radiation is elliptically polarized and now

$$\hat{\mathbf{e}}_i = \frac{\mathbf{u}_i + i\mathbf{v}_i}{\sqrt{\mathbf{u}_i^2 + \mathbf{v}_i^2}}, \quad \mathbf{u}_i \cdot \mathbf{v}_i = 0.$$

The $\hat{\mathbf{e}}_i$ now have 4 degrees of freedom and so there are $4N$ degrees of freedom in the set of $\hat{\mathbf{e}}_i$.

In total therefore, there are 7 degrees of freedom in each radiating change, and therefore 7*N* degrees of freedom available for minimization of (17), via the $f_{ij}$. From the definition (16) one sees that $|f_{ij}| \leq 1$. Further, prior to minimization, one has no reason to suppose any particular value for the phases, nor correlation with the $r_{ij}$. Therefore

$$\langle f_{ij} \rangle = 0, \quad \langle f_{ij} / r_{ij} \rangle \approx \langle f_{ij} \rangle / r_{ij} = 0. \tag{19}$$

Consequently the expected interaction energy is zero:

$$\langle \bar{\varepsilon}_{\text{int}} \rangle = 4\pi\alpha m_e \sum_{i=1}^{N} \sum_{j=i+1}^{N} \langle f_{ij} (R/r_{ij} - 1) \rangle = 0. \tag{20}$$

*3.3 Energy minimum*

A proper minimization of (20) would yield a complicated relation between the phases in $f_{ij}$ and the distances $r_{ij}$. Here we will be content with a very rough estimate of the effects of minimization, achieved by making some refinements to an initially gross approximation. We assume at first that the minimized $f_{ij}$ do not depend on the $r_{ij}$. That is, instead of (20), we will minimize the sum $\sum_{i,j} f_{ij}$. (As shown below, the expected value, averaged over all locations, of the factor $R/r_{ij} - 1$ is positive, and therefore a minimized energy requires that the $f_{ij}$ be, on average, negative.) Further, we suppose at first this is achieved by selecting just 7*N* of the $f_{ij}$ and adjusting the phases so as to set them to their minimum value, -1 leaving the remaining (vast majority) of the $f_{ij}$ retaining a zero expectation value:

$$\forall j, \forall i \in [1,7]: \langle f_{ij} \rangle \to -1, \quad \forall j, \forall i \in [8,N]: \langle f_{ij} \rangle = 0. \tag{21}$$

whereupon

$$\sum_{i,j} \langle f_{ij} \rangle = -7N. \tag{22}$$

But a more equitable procedure would be to distribute this freedom equally amongst the $f_{ij}$, in which case one would expect after minimization the same expectation (22), but now

$$\forall i,j: f_{ij} \to \langle f_{ij} \rangle = -7/N. \tag{23}$$

Eq. (23) seems a reasonable starting point for the minimum, though obviously this result could benefit from refinement. Using (23) the interaction energy (15) becomes

$$\bar{\varepsilon}_{ij} \approx -\frac{28\pi\alpha m_e}{N} (R/r_{ij} - 1). \tag{24}$$

**4. Gravitation**

*4.1 Newton's law*

Upon taking the derivative, Eq. (24) predicts that between two charges separated by distance $r_{ij}$ there will be a Newtonian gravity-like force

$$\mathbf{F}(\mathbf{r}) = -\nabla \bar{\varepsilon}(r) = \nabla \bar{\varepsilon}(r) \frac{28\pi \alpha m_e c^2}{N} \left( \frac{R}{r} - 1 \right) = -\frac{28\pi \alpha m_e c^2 R}{N} \frac{\hat{\mathbf{r}}}{r^2} \qquad (25)$$

where $c$ has been restored, and will be made explicit hereafter. Though this highly approximate model can be no more than a small step towards a fuller more comprehensive theory (see the caveats in the discussion section below), even so, Eq. (25) suggests the possibility of a novel electromagnetic basis for gravity. Let us now try to relate the coefficient in (25) to the constant of gravitation.

*4.2 Two-particle gravitational interaction*
Only the first term in parenthesis in Eq. (24) gives rise to a negative binding energy. Let us denote this by a subscript g and consider these $\bar{\varepsilon}_{ij}$ to be a continuous function of $r$:

$$\bar{\varepsilon}_g(r) \approx -\frac{28\pi \alpha m_e c^2 R}{Nr}. \qquad (26)$$

Let us now use this to make an estimate for a gravitational constant $G^*$ according to the standard formula, here applied to two charges both of mass $m_e$:

$$\bar{\varepsilon}_g(r) = -\frac{G^* m_e^2}{r}. \qquad (27)$$

Comparing (26) and (27) one infers that

$$G^* = \frac{28\pi \alpha R c^2}{N m_e}. \qquad (28)$$

Let us first examine this prediction using current estimates for $N$ and $R$ from astrophysical observation. The currently favoured view is the universe has the critical mass density satisfying

$$\rho_{\text{crit}} \equiv \frac{3H_0^2}{8\pi G} \qquad (29)$$

where $H_0$ is the Hubble constant, and $R = c/H_0$. The components of this critical density are vacuum, non-baryonic, and baryonic, whose relative contributions are denoted respectively by $\Omega_v, \Omega_{nb}, \Omega_b$. And we will need to refer to the total mass fraction $\Omega_m = \Omega_{nb} + \Omega_b$. Conventionally the contributions to the non-baryonic matter from radiation and leptons are small enough to be neglected. The baryonic matter is primarily comprised of protons. Since the number of electrons is probably very nearly equal to the number of protons, an estimate for the number of electrons is

$$N \approx \frac{4\pi R^3}{3} \frac{\Omega_b \rho_{\text{crit}}}{m_p} \approx \frac{4\pi R^3}{3} \frac{3\Omega_b c^2}{8\pi G m_p R^2} = \frac{\Omega_b R c^2}{2 G m_p}. \qquad (30)$$

With this, the ratio $R/N$ in (28) is

$$\frac{R}{N} \approx \frac{2 G m_p}{\Omega_b c^2}. \qquad (31)$$

Putting this into (28) gives

$$\frac{G^*}{G} = \frac{56\pi \alpha}{\Omega_b} \frac{m_p}{m_e}. \qquad (32)$$

The current view is that the vacuum energy makes up about 2/3 of the total critical energy density:

$$\Omega_v = \frac{2}{3}, \quad \Omega_b + \Omega_{nb} = \frac{1}{3}. \tag{33}$$

In the event that there is no missing matter then $\Omega_b \approx 1/3$ whereupon (32) gives

$$\frac{G^*}{G} = 168\pi\alpha \frac{m_p}{m_e} = 3.8 \frac{m_p}{m_e}. \tag{34}$$

At first glance the result (34) appears to be a disappointment: the estimate for the constant of gravitation is about 4000 times too high. However, the (traditional) estimate for the number of particles in the Hubble volume is based upon the assumption that nearly all the 'not-missing' mass comes from protons. Instead, in this toy universe wherein the predominant 'non-missing mass' derives entirely from electrons and positrons the calculation should probably proceed differently as follows.

It would still be reasonable to expect $\Omega_m \approx 1/3$, since, though at present lacking an explanation, this near miss (i.e. to 1) presumably derives from some fundamental – fixed – relation. Eq. (29) would be replaced with

$$\rho_{\text{crit}} \equiv \frac{3H_0^2}{8\pi G^\dagger} \tag{35}$$

where $G^\dagger$ is the constant of gravitation appropriate for a universe dominated by positrons and electrons. The number of electrons and positrons in the Hubble would then be

$$N \approx \frac{4\pi R^3}{3} \frac{\Omega_l \rho_m}{m_e} \approx \frac{4\pi R^3}{3} \frac{3\Omega_l c^2}{8\pi G^\dagger m_e R^2} = \frac{\Omega_l Rc^2}{2G^\dagger m_e} \tag{36}$$

where $\Omega_l$ is the fraction of critical mass contained in the *leptons*. The ratio $R/N$ in (28) would be

$$\frac{R}{N} \approx \frac{2G^\dagger m_e}{\Omega_l c^2}. \tag{37}$$

Putting this into (28) gives

$$\frac{G^*}{G^\dagger} = \frac{56\pi\alpha}{\Omega_l}.$$

Once again, if there is no missing mass, then $\Omega_m = \Omega_f \approx 1/3$ and then one has

$$\frac{G^*}{G^\dagger} = 168\pi\alpha = 3.8. \tag{38}$$

Given the approximations employed in deriving (28) this agreement within a factor of 4 is quite satisfactory.

*4.3 Total gravitational interaction energy*
From (17) and (24), the total interaction energy is

$$\bar{\varepsilon}_{\text{int}} = -\frac{28\pi\alpha m_e}{N} \sum_{i=1}^{N} \sum_{j=i+1}^{N} (R/r_{ij} - 1). \tag{39}$$

For a uniform distribution of matter one has

$$\sum_{i=1}^{N}\sum_{j=i+1}^{N} R/r_{ij} \approx \left(N^2/2\right)\left\langle R/r_{ij}\right\rangle = \left(N^2/2\right)R\int_0^R dr\, r^2\left(1/r\right)\Big/\int_0^R dr\, r^2 = \frac{3}{4}N^2 \tag{40}$$

and therefore

$$\sum_{i=1}^{N}\sum_{j=i+1}^{N}\left(R/r_{ij}-1\right) \approx N^2/4. \tag{41}$$

Putting this into (39) gives

$$\overline{\varepsilon}_{\text{int}} \approx -7\pi\alpha N m_e = -\kappa\overline{\varepsilon}_m \tag{42}$$

where

$$\kappa \equiv 7\pi\alpha = 0.16 \tag{43}$$

and where $\overline{\varepsilon}_m = N m_e$ is the average mechanical energy. It is concluded from Eq. (42) that minimization through adjustment of phases has led to an overall interaction energy that is within one order of the total mass energy (of this positron-electron universe).

*4.4 Role of the Virial Theorem*
Though leading to a plausible result, this minimization procedure leaves much to be desired. In particular the following three areas could be improved:

i) The averaging process (40) takes some liberty with shifting the origin of the $r_i$ and $r_j$ without due respect for a consistent geometry.

ii) The numerical coefficient in (42) was obtained by a rough estimation of the $f_{ij}$. But a proper calculation requires that the micro-dynamics of the circulating charges (i.e. the zitterbewegung) be computed in response to the direct-action fields so that the whole system is self-consistent.

iii) One could argue that the maximum impact of the 7$N$ degrees of freedom would have been achieved if, instead of being distributed as a uniform offset amongst the $f_{ij}$, they were targeted on nearest pair interactions, i.e. where the $r_{ij}$ were minimal.

There is no doubt then that the result (42) could be refined. Whereas items i) and ii) are of uncertain impact on the interaction energy, item iii) will cause it to increase in magnitude. There is an upper limit however, to the magnitude of the interaction energy. The relativistic virial theorem [31] applies here without approximation because the system is completely closed. (In the direct action version of EM, there are no radiation fields.) Continuing with the notation introduced in (1), the result is that the total system energy, i.e., the inertial plus interaction energy, is

$$\varepsilon_{\text{tot}} = m_{\text{bare}}\sum_{i=1}^{N}\overline{\sqrt{1-\mathbf{v}_i^2}}, \tag{44}$$

where the bar denotes a time average. In accord with the classical zitterbewegung conjecture, the observed mass $m_e$ is a time-averaged quantity whose value is allegedly the same for all sources:

$$m_e = m_{\text{bare}}\overline{1/\sqrt{1-\mathbf{v}_i^2}} \tag{45}$$

where $m_{\text{bare}} \to 0$, $|\mathbf{v}_i| \to 1$ as the quantity on the right of (45) tends to the finite observed value. One infers from (44) that the time-averaged interaction energy is

$$\bar{\varepsilon}_{\text{int}} = -m_{\text{bare}} \sum_{i=1}^{N} \overline{\left( \frac{1}{\sqrt{1-\mathbf{v}_i^2}} - \sqrt{1-\mathbf{v}_i^2} \right)} = -m_{\text{bare}} \sum_{i=1}^{N} \overline{\frac{\mathbf{v}_i^2}{\sqrt{1-\mathbf{v}_i^2}}}. \tag{46}$$

The ratio of interaction energy to the zitterbewegung-boosted mechanical mass energy, is

$$\bar{\varepsilon}_{\text{int}} / \bar{\varepsilon}_{\text{m}} = -\sum_{i=1}^{N} \overline{\frac{\mathbf{v}_i^2}{\sqrt{1-\mathbf{v}_i^2}}} \bigg/ \sum_{i=1}^{N} \overline{\frac{1}{\sqrt{1-\mathbf{v}_i^2}}} \geq -1. \tag{47}$$

The equality holds if the zitterbewegung motion is at the speed of light, at which point

$$\varepsilon_{\text{tot}} = \bar{\varepsilon}_{\text{int}} + \bar{\varepsilon}_{\text{m}} = 0, \tag{48}$$

replacing the result (42). This is in accord with the prevailing view in cosmology, wherein one may interpret the critical density condition $\Omega = 1$ as zero total system energy. It should be noted however that (48) implies a revision of the $f_{ij}$, which may have some impact on the result (26) and therefore on (38). The final result of that revision must be that (43) should be replaced with $\kappa = 1$.

*4.5 Ambient interaction*

With reference to (24) one sees that associated with each charge pair there is a net positive energy, independent of the spatial distribution of the sources, equal to

$$\bar{\varepsilon}_{ij}(\infty) \approx 28\pi\alpha m_{\text{e}} / N, \tag{49}$$

and therefore a net positive energy $\bar{\varepsilon}_{\text{a}}$, say,

$$\bar{\varepsilon}_{\text{a}} \equiv \sum_{i=1}^{N} \sum_{j=i+1}^{N} \bar{\varepsilon}_{ij}(\infty) \approx 14\pi\alpha N m_{\text{e}}. \tag{50}$$

Given the very approximate minimization procedure above it is difficult to know how much significance to give to this term. It is conceivable that determination of the $f_{ij}$ through a more accurate minimization of (15) would leave no interaction terms that were completely independent of particle separation. However, if it exists, it is not clear that this energy could be absorbed into a definition of the inertial masses because it is a result of interactions and so is non-local in character. On the other hand, this energy does not depend on particle separations, and therefore does not generate a force. Let us refer to the sum over all pairs of these contributions as the 'ambient' energy, $\bar{\varepsilon}_{\text{a}}$. Then the total interaction energy given in (42) can be written $\bar{\varepsilon}_{\text{int}} = \bar{\varepsilon}_{\text{a}} + \bar{\varepsilon}_{\text{g}}$ where

$$\bar{\varepsilon}_{\text{a}} \approx 2\kappa\bar{\varepsilon}_{\text{m}}, \quad \bar{\varepsilon}_{\text{g}} \approx -3\kappa\bar{\varepsilon}_{\text{m}} \tag{51}$$

in accord with the averaging process in (40) and (41).

If, as before, we invoke the virial theorem to correct the coefficient $\kappa$, setting $\kappa = 1$. Then one has

$$\bar{\varepsilon}_{\text{tot}} = \bar{\varepsilon}_{\text{a}} + \bar{\varepsilon}_{\text{m}} + \bar{\varepsilon}_{\text{g}} = 0, \quad \bar{\varepsilon}_{\text{a}} : \bar{\varepsilon}_{\text{m}} : \bar{\varepsilon}_{\text{g}} = 2:1:-3. \tag{52}$$

It is noteworthy that the ambient and matter terms contribute to the positive energy in the ratio 2:1 (67%, 33%), in accord with currently popular estimates for the dark energy to matter ratio. This appears to suggest that the ambient energy should be identified with dark energy, lending additional support to the model under discussion. Due to the very approximate character of the above calculations, however, we are a long way from being able to assert this connection with any confidence.

## 5. Discussion

The model presented here succeeds in deriving a Newtonian gravitation-like force from a purely electromagnetic direct-action theory. The picture it paints is of a binding energy which is a consequence of maximal phase correlation (or anti-correlation) of the sources. The resulting force can be regarded as a strong version of the van der Waals force: whereas the ordinary van der Waals force is derived from phase coherence of *secondary* radiation in response to completely incoherent primary radiation (the ZPF), the superior ($1/r^2$) strength of the force of gravity can be attributed to the primary role of relatively coherent direct-action fields. Also, when averaged over the cosmological scale, the model predicts an ambient positive energy of magnitude exactly twice that of inert matter, inviting comparison with dark energy.

There are some major deficiencies in the current presentation of this model, however, the most important of which are:

i) The phase factors $f_{ij}$ have been computed only approximately, as evinced by comparison of the total minimized energy with the total system energy given by the virial theorem.
ii) The matter distribution is assumed to be uniform.
iii) Lorentz Invariance of the approximately monochromatic background classical field has not been established.
iv) Use has been made of the steady-state cosmology.
v) Self-consistency of the predominantly circular motion of charges has not yet been established in detail beyond the energetic level addressed by the virial theorem.
vi) No indication has been given of how, at elevated temperatures, exclusively retarded radiation can emerge from this implementation of the direct-action theory.
vii) The model universe acknowledges only electrons and positrons, with no indication of how protons should be accommodated.
viii) The ambient energy has not been shown to be Lorentz invariant, and therefore its relation to dark energy is still uncertain.
ix) The Newtonian law has been derived assuming the masses are static; the relativistic domain has not been considered.

Given the present level of approximation therefore, it is too early to come to a firm conclusion on the validity of this approach though arguably the results suggest that further investigation is warranted.

**6. Acknowledgements**
I am very grateful to Harold Puthoff for introducing me to stochastic electrodynamics. Though in the end SED does not succeed in providing an EM basis for gravity, it was only in coming to understand what was missing that the attractions of an alternative approach based on direct-action became apparent.